# Resonant Interaction Between a Weak Gravitational Wave and a Microwave Beam in the Double Polarized States Through a Static Magnetic Field *

Fang-Yu LI[+],    Nan YANG[++]

+ Department of Physics, Chongqing University, Chongqing 400044,China

E-mail: fangyuli@cqu.edu.cn

+ +Department of Physics, Chongqing University, Chongqing 400044,China

E-mail: nenyan@hotmail.com

We investigate the resonant interaction to the weak gravitational waves in a coupling electromagnetic system, which consists of a Gaussian beam with the double polarized transverse electric modes, a static magnetic field and the fractal membranes. We find that under the syncroresonance condition a high-frequency GW (HFGW) of $h = 10^{-30}$, $\nu_g = 3\text{GHz}$ may produce the perturbative photon flux (PPF) of $2.15 \times 10 \text{s}^{-1}$ in a surface of $10^{-2} \text{m}^2$. The PPF can be pumped out from the background photon fluxes and one might obtain the amplified signal photon flux of $2.15 \times 10^4 \text{s}^{-1}$ by cascade fractal membranes. It appears to be worthwhile to study this effect for the detection of the high-frequency relic GWs in quintessential inflationary models and the HFGWs expected by possible laboratory schemes.

*Supported by the National Natural Science Foundation of China under Grant No. 10175096, the National Basic Research Program of China under Grant No.2003CB716300 and Natural Science Foundation of Chongqing.

One important expectation [1,2] of quintessential inflationary models is that the maximal signal of the relic gravitational waves (GWs) in the models may be firmly localized in the GHz band, the corresponding dimensionless amplitude of the relic GWs can reach up to roughly $10^{-30}$. Although the energy density of the relic gravitons in the region is almost eight orders of magnitude larger than in ordinary inflationary models, they have not yet been detected. Moreover, fast development of a series of new technology (nanotechnology, ultra-fast science, high-temperature superconductors, ultra-strong field physics, etc.) offered new hopes for laboratory generation of the high-frequency GWs (HFGWs) [3-5], but the orders of the amplitude of the HFGWs expected by possible laboratory schemes in the GHz band would be only $10^{-30} - 10^{-31}$ or less.

In this letter we shall show that under the syncroresonance condition the perturbative photon flux induced by a continuous monochromatic HFGW of $h = 10^{-30}$, $v_g = 3\text{GHz}$ can reach up to $2.15 \times 10 \text{s}^{-1}$ at a surface of $10^{-2} \text{m}^2$ in a coupling system between the Gaussian-type microwave beam and the static magnetic field, and the perturbative photon flux can be pumped out from the background photon fluxes by the special fractal membranes [6-8]. Using the cascade fractal membranes or crystal channels effect [9] the signal photon flux may increase 2-3 orders of magnitude, while the instantaneous value of the perturbative photon flux produced by the high-frequency relic GW of $h = 10^{-30}$ in the GHz band, might reach up to the same orders of magnitude. Thus, such effects are possible to provide a new displaying way for the HFGWs.

It is well know that in flat spacetime there are many different kinds of forms of the Gaussian-type beams. In fact, they are the wave beam solutions of the electrodynamical equations in flat spacetime, while the Gaussian beam of fundamental frequency mode has simplest expression [10], i.e.,

$$\psi = \frac{\psi_0}{\sqrt{1+(z/f)^2}} \exp(-\frac{r^2}{W^2}) \exp\left\{i[(k_e z - \omega_e t) - \tan^{-1}\frac{z}{f} + \frac{k_e r^2}{2R} + \delta]\right\}, \qquad (1)$$

where $r^2 = x^2 + y^2$, $k_e = 2\pi/\lambda_e$, $f = \pi W_0^2/\lambda_e$, $W = W_0[1+(z/f)^2]^{1/2}$, $R = z + f^2/z$, $\psi_0$ is the amplitude of the electric (or magnetic) field of the Gaussian beam, $W_0$ is the minimum spot radius, $\omega_e$ is the angular frequency, $\delta$ is an arbitrary phase factor.

Unlike Refs. [11,12], here we choose the Gaussian beam with both the transverse polarized electric modes (TE), and we adopt the coupling system between the Gaussian beam, the fractal membranes and the static magnetic field. The static magnetic field is pointed along the y-axis and is localized in the region $-l/2 \leq z \leq l/2$. Although then the form of the perturbative photon fluxes will be more complicated, they have more realistic observable meaning. Setting $\psi = \psi_x = \tilde{E}_x^{(0)}$ in Eq. (1), using divergenceless condition $\nabla \cdot \mathbf{E} = \frac{\partial \psi_x}{\partial x} + \frac{\partial \psi_y}{\partial y} = 0$ and $\tilde{\mathbf{B}}^{(0)} = -\frac{i}{\omega_e} \nabla \times \tilde{\mathbf{E}}^{(0)}$ (we use MKS units), we have

$$\tilde{E}_x^{(0)} = \psi = \psi_x, \quad \tilde{E}_y^{(0)} = \psi_y = -\int \frac{\partial \psi_x}{\partial x} dy = 2x(\frac{1}{W^2} - i\frac{k_e}{2R})\int \psi_x dy, \quad \tilde{E}_z^{(0)} = 0,$$

(2)

$$\tilde{B}_x^{(0)} = \frac{i}{\omega_e} \frac{\partial \psi_y}{\partial z}, \quad \tilde{B}_y^{(0)} = -\frac{i}{\omega_e} \frac{\partial \psi_x}{\partial z}, \quad \tilde{B}_z^{(0)} = \frac{i}{\omega_e}(\frac{\partial \psi_x}{\partial y} - \frac{\partial \psi_y}{\partial x}),$$

besides,

$$B^{(0)} = \hat{B}^{(0)} = \begin{cases} \hat{B}_y^{(0)} & (-l/2 \leq z \leq l/2), \\ 0 & (z \leq -l/2 \text{ and } z \geq l/2), \end{cases}$$

(3)

where the superscript 0 denotes the background electromagnetic (EM) fields, the notations ~ and ^ stand the time-dependent and static EM fields, respectively. For the high-frequency EM energy flux densities (or in quantum language: photon flux densities), only nonvanishing average values of these with respect to time have an observable effect. From Eqs.(1) and (2), one finds

$$n_x^{(0)} = \frac{1}{\hbar\omega_e}\left\langle \frac{1}{\mu_0}\tilde{E}_y^{(0)}\tilde{B}_z^{(0)}\right\rangle = \frac{1}{2\mu_0\hbar\omega_e}\text{Re}\left\{\psi_y^*\left[\frac{i}{\omega_e}\left(\frac{\partial\psi_x}{\partial y}-\frac{\partial\psi_y}{\partial x}\right)\right]\right\}$$

$$= f_x^{(0)}\exp\left(-\frac{2r^2}{W^2}\right),$$

$$n_y^{(0)} = -\frac{1}{\hbar\omega_e}\left\langle \frac{1}{\mu_0}\tilde{E}_x^{(0)}\tilde{B}_z^{(0)}\right\rangle = \frac{1}{2\mu_0\hbar\omega_e}\text{Re}\left\{\psi_x^*\left[\frac{i}{\omega_e}\left(\frac{\partial\psi_y}{\partial x}-\frac{\partial\psi_x}{\partial y}\right)\right]\right\} \quad (4)$$

$$= f_y^{(0)}\exp\left(-\frac{2r^2}{W^2}\right),$$

$$n_z^{(0)} = \frac{1}{\hbar\omega_e}\text{Re}\left\langle \frac{1}{\mu_0}(\tilde{E}_x^{(0)}\tilde{B}_y^{(0)}) - \frac{1}{\mu_0}(\tilde{E}_y^{(0)}\tilde{B}_x^{(0)})\right\rangle$$

$$= -\frac{1}{2\mu_0\hbar\omega_e}\text{Re}\left\{\psi_x^*(\frac{i}{\omega_e}\frac{\partial\psi_x}{\partial z}) + \psi_y^*(\frac{i}{\omega_e}\frac{\partial\psi_y}{\partial z})\right\} = f_z^{(0)}\exp\left(-\frac{2r^2}{W^2}\right),$$

where $n_x^{(0)}$ $n_y^{(0)}$ and $n_z^{(0)}$ represent the average values of the background photon flux densities propagating along the x-, y- and z-axes, respectively, the angular brackets denote the average over time, $f_x^{(0)}$, $f_y^{(0)}$ and $f_z^{(0)}$ are the functions of $\psi_0, W_0, \omega_e, r$ and $z$. Because of the non-vanishing $n_x^{(0)}$ and $n_y^{(0)}$, the Gaussian beam will be asymptotically spread as $|z|$ increases.

For a circular polarized weak monochromatic plane GW propagating along the z-axis, the line element takes the form

$$ds^2 = -c^2dt^2 + (1+h_\oplus)dx^2 + (1-h_\oplus)dy^2 + 2h_\otimes dxdy + dz^2$$

$$= -c^2dt^2 + \{1 + A_\oplus\exp[i(k_g z - \omega_g t)]\}dx^2 + \{1 - A_\oplus\exp[i(k_g z - \omega_g t)]\}dy^2 \quad (5)$$

$$+ i2A_\otimes\exp[i(k_g z - \omega_g t)]dxdy + dz^2,$$

where $A_\oplus$ and $A_\otimes$ in Eq.(5) are the constant amplitudes. For the relic GWs, they are actually the instantaneous values at some instant, which contain the cosmology scale factor [13]. Since their typical orders are about $10^{-30}$ $10^{-31}$ in the GHz band [1,2], these weak gravitational processes in the laboratory frame of reference of the Earth can be described as the perturbations to the Euclidian space. Of course, then the time and frequency parameters should be the laboratory time and laboratory frequency.

Using the electrodynamical equations in curved spacetime, the generic form of the perturbative EM fields produced by the direct interaction of the GW with the static field $\hat{B}_y^{(0)}$ can be given by [11]

$$\tilde{F}_{\mu\nu}^{(1)} = a(z)\exp[i(k_g z - \omega_g t)] + b(z)[i(k_g z + \omega_g t)], \qquad (6)$$

where $\tilde{F}_{\mu\nu}^{(1)}$ represent $\tilde{E}_x^{(1)}, \tilde{E}_y^{(1)}$ and $\tilde{B}_x^{(1)}, \tilde{B}_y^{(1)}$, and $\tilde{E}_y^{(1)} = \tilde{B}_y^{(1)} = 0$. The orders of $\tilde{E}_x^{(1)}, \tilde{E}_y^{(1)}$ and $\tilde{B}_x^{(1)}, \tilde{B}_y^{(1)}$ within the spot radius are approximately $A\hat{B}_y^{(0)}c$ and $A\hat{B}_y^{(0)}$ [11,12], respectively, here we assume that $A_\oplus = A_\otimes = A$. The coherent syncroresonance between the perturbative EM fields, Eq. (6), and the Gaussian beam, Eqs. (1) and (2), can be expressed as the following first order perturbative photon flux (PPF) densities:

$$\begin{aligned}
n_x^{(1)} &= \frac{1}{\hbar\omega_e}\left\langle S_x^{(1)}\right\rangle_{\omega_e=\omega_g} = \frac{1}{\hbar\omega_e}\left\langle \frac{1}{\mu_0}\tilde{E}_y^{(1)}\tilde{B}_z^{(0)}\right\rangle_{\omega_e=\omega_g} \\
&= \frac{1}{2\mu_0\hbar\omega_e}\mathrm{Re}\left\{\tilde{E}_y^{(1)*}\left[\frac{i}{\omega_e}(\frac{\partial\psi_x}{\partial y}-\frac{\partial\psi_y}{\partial x})\right]\right\}_{\omega_e=\omega_g} = f_x^{(1)}\exp(-\frac{r^2}{W^2}),
\end{aligned} \qquad (7)$$

$$\begin{aligned}
n_y^{(1)} &= \frac{1}{\hbar\omega_e}\left\langle S_y^{(1)}\right\rangle_{\omega_e=\omega_g} = -\frac{1}{\hbar\omega_e}\left\langle \frac{1}{\mu_0}\tilde{E}_x^{(1)}\tilde{B}_z^{(0)}\right\rangle_{\omega_e=\omega_g} \\
&= \frac{1}{2\mu_0\hbar\omega_e}\mathrm{Re}\left\{\tilde{E}_x^{(1)*}\left[\frac{i}{\omega_e}(\frac{\partial\psi_x}{\partial y}-\frac{\partial\psi_y}{\partial x})\right]\right\}_{\omega_e=\omega_g} = f_y^{(1)}\exp(-\frac{r^2}{W^2}),
\end{aligned} \qquad (8)$$

$$\begin{aligned}
n_z^{(1)} &= \frac{1}{\hbar\omega_e}\left\langle S_z^{(1)}\right\rangle_{\omega_e=\omega_g} = \frac{1}{\hbar\omega_e}\left[\left\langle\frac{1}{\mu_0}\tilde{E}_x^{(0)}\tilde{B}_y^{(1)}\right\rangle_{\omega_e=\omega_g} + \left\langle\frac{1}{\mu_0}\tilde{E}_x^{(1)}\tilde{B}_y^{(0)}\right\rangle_{\omega_e=\omega_g}\right. \\
&\quad\left. -\left\langle\frac{1}{\mu_0}\tilde{E}_y^{(0)}\tilde{B}_x^{(1)}\right\rangle_{\omega_e=\omega_g} - \left\langle\frac{1}{\mu_0}\tilde{E}_y^{(1)}\tilde{B}_x^{(0)}\right\rangle_{\omega_e=\omega_g}\right] \\
&= \frac{1}{2\mu_0\hbar\omega_e}\left\{\mathrm{Re}\left(\tilde{B}_y^{(1)*}\psi_x\right)_{\omega_e=\omega_g} - \mathrm{Re}\left(\tilde{B}_x^{(1)*}\psi_y\right)_{\omega_e=\omega_g}\right. \\
&\quad\left. -\mathrm{Re}\left[\tilde{E}_x^{(1)*}\left(\frac{i}{\omega_e}\frac{\partial\psi_x}{\partial z}\right)\right]_{\omega_e=\omega_g} - \mathrm{Re}\left[\tilde{E}_y^{(1)*}\left(\frac{i}{\omega_e}\frac{\partial\psi_y}{\partial z}\right)\right]_{\omega_e=\omega_g}\right\} \\
&= f_z^{(1)}\exp(-\frac{r^2}{W^2}),
\end{aligned} \qquad (9)$$

where $n_x^{(1)}, n_y^{(1)}$ and $n_z^{(1)}$ are the average values of the first-order PPF densities along the $x-$, $y-$ and $z-$ axes, respectively. It is easily seen from Eqs. (2) and (4) that $\tilde{E}_y^{(0)}\big|_{x=0}=0$, thus $n_x^{(0)}\big|_{x=0}=0$. Unlike $n_x^{(0)}\big|_{x=0}=0$, it can be shown from Eq. (7) that $n_x^{(1)}\big|_{x=0}\neq 0$, and numerical calculation shows that $n_x^{(1)}\big|_{x=0}$ has maximum, this means that the photon flux passing through the yz-plan will be the pure first-order PPF, this is satisfactory. Therefore, any photon measured from the yz-plan (i.e. the plan x=0) will be a signal of the EM perturbation produced by the GW. Nevertheless, for the regions of $x\neq 0$, we have $n_x^{(0)}\neq 0$. At first sight, $n_x^{(1)}$ will be swamped by the background $n_x^{(0)}$ so that $n_x^{(1)}$ has no observable effect in the regions. However, it will be shown that $n_x^{(1)}$ and $n_x^{(0)}$ propagate along opposite directions in some local regions (see below), so that $n_x^{(1)}$ (signal), in principle, can be observed. The total PPF passing through a certain "typical receiving surface" at the yz-plane will be

$$N_x^{(1)} = \iint_{\Delta s} n_x^{(1)}\bigg|_{x=0} dydz. \qquad (10)$$

Notice that $N_x^{(1)}$ is unique non-vanishing photon flax passing through the surface.

The outgoing (and imploding) property of $n_x^{(0)}$ in the region of $z>0$ (and $z<0$) (this is just typical property of the Gaussian beams [10]), the continuity of $n_x^{(1)}$ at the $yz$-plane and the anti-symmetric relation of the propagating directions of $n_x^{(1)}$ between regions of $y>0$ and $y<0$ show that $n_x^{(1)}$ and $n_x^{(0)}$ propagate along opposite directions in the regions of 1st ($x>0$, $y>0$, $z>0$), 3rd ($x<0$, $y<0$, $z>0$), 6th ($x<0$, $y>0$, $z<0$) and 8th ($x>0$, $y<0$, $z<0$) octants, while they have the same propagating directions in the regions of 2nd, 4th, 5th and 7th octants. Thus for the effect of the GW, we are interested in the former but not in the latter.

By using Eqs. (1), (2), (6) and (7), we get the concrete form of $n_x^{(1)}$ in the region $-l/2 \leq z \leq l/2$ as follows

$$n_x^{(1)} = -\frac{1}{\hbar\omega_e}\left\{\frac{A\hat{B}_y^{(0)}\psi_0 k_g y(z+l/2)}{4\mu_0[1+(z/f)^2](z+f^2/z)}\sin(\frac{k_g r^2}{2R}-arctg\frac{z}{f})\exp(-\frac{r^2}{W^2})\right.$$

$$+\frac{A\hat{B}_y^{(0)}\psi_0 k_g F(y)(z+l/2)}{4\mu_0 R[1+(z/f)^2]^{1/2}}(1-\frac{4x^2}{W^2})\sin(\frac{k_g r^2}{2R}-arctg\frac{z}{f})\exp(-\frac{x^2}{W^2})$$

$$+\frac{A\hat{B}_y^{(0)}\psi_0 y(z+l/2)}{4\mu_0 W_0^2[1+(z/f)^2]^{3/2}}\cos(\frac{k_g r^2}{2R}-arctg\frac{z}{f})\exp(-\frac{r^2}{W^2})$$

$$\left.+\frac{A\hat{B}_y^{(0)}\psi_0 F(y)(z+l/2)}{4\mu_0 W_0^2[1+(z/f)^2]^{1/2}}[1+(\frac{k_g}{R}-\frac{2}{W^2})x^2]\cos(\frac{k_g r^2}{2R}-arctg\frac{z}{f})\exp(-\frac{x^2}{W^2})\right\},$$

(11)

where

$$F(y) = \int\exp(-\frac{y^2}{W^2})\cos(\frac{k_g}{2R}y^2)dy,$$

(12)

is quasi-probability integral.

In our EM system, all parameters are chosen as realizable values in the present experiments: (1) P=10W, the power of the Gaussian beam. (2) $\hat{B}_y^{(0)} = 3T$, the strength of the background static magnetic field, (3) $W_0 = 0.1m$, the spot radius of the Gaussian beam (4) $W_0 \leq y \leq 2W_0, 0 \leq z \leq l/2$ ($W_0 = 0.1m$, $l = 0.3m$), the integration region $\Delta s$ in Eq (13), i.e., $\Delta s = 1.5\times 10^{-2} m^2$. Obviously then "the receiving surface" has already moved to the region outside the spot radius $W_0$, it has a more realistic meaning. (5) $A = 10^{-30}$, $\nu_g = 3GHz$, they are not only typical orders of the high-frequency relic GWs in the quintessential inflationary models [1,2], but also typical orders predicted by possible laboratory HFGW schemes [3,5]. Using Eqs.(1), (2), (6), (7), (10),(11) and the above parameters, we obtain $N_x^{(1)} = 2.15\times 10 s^{-1}$. For the continuous GW, it corresponds $7.74\times 10^4 h^{-1}$.

It should be pointed out that because of random property of propagating directions, frequencies and amplitudes of the relic GW's [4,13], detection of the relic GW's will be more difficult than that of the monochromatic continuous GW's. However, a GW propagating along an arbitrarily direction can always be split into three components, which propagate along $x-$, $y-$ and $z-$ directions, respectively. And as we have shown above that only the GW, which satisfies the resonant condition ($\omega_g = \omega_e$) and propagates along positive direction of the z-axis, can generate optimal resonant response. It can be shown that for the GW's propagating along x-, $y$-axises and negative direction of the $z$-axis, even if $\omega_g = \omega_e$, the effective PPFs produced by them will be much less than that generated by the GW propagating along the positive direction of the $z$-axis. Consequently, our EM system has very strong selection capability to the relic GW's. While the relic GW's have not yet been detected, we can be sure that the Earth is bathed in sea of the relic GW's. Therefore, the high-frequency relic GW's satisfying above "frequency and direction resonance" would be selectable and measurable in seconds at least.

It is remarkable that in recent years the new-type fractal membranes have been successfully developed (see e.g. Ref.[6-8]), and first, the fractal membranes can provide nearly total reflection for the EM waves (photon fluxes) with certain frequencies in the GHz band, at the same time, they can provide nearly total transmission for the photon fluxes with other frequencies in the GHz band, and one can get amplification of ten times at each reflection or transmission. Second, the photon fluxes reflected and transmitted by the fractal membranes can keep their strength invariant within distance of 1~2 meters. Third, such frequencies can be regulated in the GHz band.

Since $N_x^{(1)}$ (signal) and $N_x^{(0)}$ (background) have opposite propagating directions in the local regions (e.g. in the first octant, $N_x^{(1)}$ and $N_x^{(0)}$ propagate along the negative and positive directions of the $x$-axis, respectively), using the fractal membrane with the normal direction parallel to the positive direction of the $x$-axis, it will reflect only $N_x^{(1)}$ and not $N_x^{(0)}$. Once $N_x^{(1)}$ is reflected, $N_x^{(1)}$ and $N_x^{(0)}$ will have the same propagating direction. However, because after $N_x^{(1)}$ is reflected, it will get amplification of 10 times and keep its strength invariant within one meter distance at least, while $N_x^{(0)}$ will be decay as $\exp(-2r^2/W^2)$ [see, Eq. (4)], then ratio $N_x^{(1)}/N_x^{(0)}$ would be larger than unity in the whole region of $0.68\mathrm{m} \leq x \leq 1\mathrm{m}$ ($x$ is the distance to the fractal membrane). If $N_x^{(1)}$ is reflected at one time (the single fractal membrane), one can get the amplified signal photon flux of $2.15 \times 10^2 \mathrm{s}^{-1}$ in a surface of $10^{-2}\mathrm{m}^2$, and it can still keep the above value at $x = 0.79\mathrm{m}$ while where $N_x^{(0)}$ will be reduced to $\sim 10^{-6}\mathrm{s}^{-1}$, although $N_x^{(0)} = N_x^{(0)}{}_{\max} \gg N_x^{(1)}$ at $x = \sqrt{2}/2 W_0 \approx 0.07\mathrm{m}$. If $N_x^{(1)}$ is reflected or transmitted at 3 times (the cascade fractal membranes), then the terminal receiver would get the signal photon flux of $2.15 \times 10^4 \text{s}^{-1}$. However, the increment of the number of the fractal membranes will bring new noise sources (the thermal noise and shot noise). In order to suppress the thermal noise, the whole system should be cooled down to $KT < \hbar \omega_e$ which corresponds to $T < 0.1K$, suppressing the shot noise is also difficult, but worthwhile and not impossible. In fact, since "random motion" of the thermal photons and the highly directional propagated property of the PPF, the requirements of distinguishing such two kinds of photons can be further relaxed. As for the $N_y$ and $N_z$, since $N_x$, $N_y$ and $N_z$ are orthogonal each other, even if the position of the fractal membrane occurs a small deviation, $N_x$, $N_y$ and $N_z$ will be orthogonal reflected to different directions and regions. This means that such fractal membrane can be made an "equivalent light splitter" for the signal and the background.

For the possible external EM noise sources, using a Faraday cage or shielding covers made from such fractal membranes would be very helpful. Moreover, since $n_x^{(1)} \propto \hat{B}_y^{(0)}$ [see, Eq. (11)], increasing the background static magnetic field may be a better way (e.g., using crystal channel effect [9] or other means [14], it is possible to get a much stronger static fields than 3 Tesla). In this case $n_x^{(1)}$ would be increased 2-3 orders, and the number of the background real photons does not change. These questions will be considered elsewhere.

One of the authors (Li F Y) would like to thank Dr. Baker for very useful discussions and suggestions.